\documentclass[fp,twocolumn,seceq,dvipdfmx]{jpsj3}
\usepackage{graphicx}
\def\vv#1{\mib #1}
\def\dfrac#1#2{{\displaystyle\frac{#1}{#2}}}

\newcommand{\ket}[1]{\left\vert {#1} \right\rangle}

\newcommand{\aver}[1]{\left\langle {#1} \right\rangle}
\newcommand{\so}[2]{\vv{S}_{#1,#2}}
\newcommand{\snpo}[1]{\vv{n}'(x_{#1})}
\newcommand{\sre}[3]{A_{#2}\vv{l}(x_{#1})#3S\sqrt{1-A^2_{#2}\dfrac{\vv{l}(x_{#1})^2}{S^2}}\vv{n}(x_{#1})}
\title
{
Partial Ferrimagnetism in $S=1/2$ Heisenberg Ladders with a Ferromagnetic Leg, an Antiferromagnetic Leg, and Antiferromagnetic Rungs
}

\author
{
Kazutaka Sekiguchi\thanks{Present address: Akikusa Gakuen High School, Sayama, Saitama 350-1312,  Japan } and Kazuo Hida\thanks{E-mail: hida@mail.saitama-u.ac.jp}
}

\inst
{Division of Material Science, Graduate School of Science and Engineering, \\ Saitama University, Saitama, Saitama 338-8570, Japan
}

\recdate
{ May 13, 2017; accepted June 13, 2017; published online July 19, 2017}

\abst
{
Ground-state and finite-temperature properties of $S=1/2$ Heisenberg ladders with a ferromagnetic leg, an antiferromagnetic leg, and antiferromagnetic rungs are studied. It is shown that a partial ferrimagnetic phase extends over a wide parameter range in the ground state. The numerical results are supported by an analytical calculation based on a mapping onto the nonlinear $\sigma$ model and a perturbation calculation from the strong-rung limit. It is shown that the partial ferrimagnetic state is a spontaneously magnetized Tomonaga--Luttinger liquid with incommensurate magnetic correlation, which is confirmed by a DMRG calculation. The finite-temperature magnetic susceptibility is calculated using the thermal pure quantum state method. It is suggested that the susceptibility diverges as $T^{-2}$ in the ferrimagnetic phases as in the case of ferromagnetic Heisenberg chains.
}

\begin{document}

\sloppy

\maketitle

\section{Introduction}

Ferrimagnetism in one-dimensional quantum magnets has been attracting broad interest in condensed matter physics. Conventional ferrimagnetism in unfrustrated spin chains can be understood on the basis of the Lieb--Mattis (LM) theorem\cite{lieb-mattis}, for which the spontaneous magnetization is quantized to the values expected from the LM theorem.\cite{kuramoto,yamamoto} This type of ferrimagnetism is called LM ferrimagnetism. For weak frustration,  LM ferrimagnetism often remains stable. Another type of quantum ferrimagnetism induced by frustration for which the spontaneous magnetization varies continuously with the strength of frustration is called partial ferrimagnetism.\cite{sachdev,tt,ir,ym,khferri,filho,htsdec,shimo1,shimo2,tonegawa} In this case, the spontaneous magnetization is not quantized to a specific value.  In many numerical examples,\cite{ym,khferri,filho,shimo1,shimo2,htsdec} partial ferrimagnetism is accompanied by an incommensurate quasi-long-range modulation of the magnetization. Recently, an analytical approach using the nonlinear $\sigma$ model has been proposed to understand the partial ferrimagnetism of this kind.\cite{furuya} It is proposed that this phase can be characterized as a spontaneously magnetized Tomonaga--Luttinger  liquid (SMTLL).

In the present work, we investigate the partial ferrimagnetism in $S=1/2$ Heisenberg ladders with a ferromagnetic leg, an antiferromagnetic leg, and antiferromagnetic rungs. In the absence of rung interactions, the system decouples to a spin-1/2 antiferromagnetic chain and a spin-1/2 ferromagnetic chain. Hence, the ground state has magnetization $M=L/2$, where $L$ is the length along the legs. On the other hand, in the strong-rung limit, two spins on each rung form a singlet dimer and the ground state is nonmagnetic with $M=0$. This ground state is called the rung-dimer state. Hence, it is plausible that a partial ferrimagnetic ground state is realized in an appropriate range of the rung strength. 

This paper is organized as follows. The Hamiltonian is introduced in Sect. 2. The ground-state phase diagram is investigated numerically and analytically in Sect. 3. The finite-temperature magnetic susceptibility is numerically estimated in Sect. 4 using the canonical thermal pure quantum state (cTPQ) method. The last section is devoted to a summary and discussion.

\section{Hamiltonian}
\begin{figure}[htbp]
\centering{\includegraphics[width=60mm]{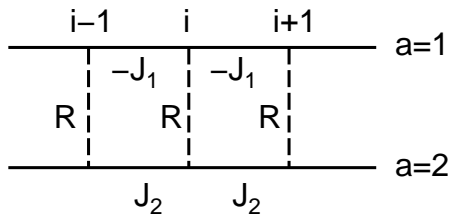}}
\caption{Lattice structure of the present model.}
\label{fig:lattice}
\end{figure}

We consider the $S=1/2$ Heisenberg ladders described by the Hamiltonian
\begin{align}
{\cal H}&=-J_1\sum_{i=1}^{L}\vv{S}_{i,1}\cdot\vv{S}_{i+1,1} +J_2\sum_{i=1}^{L}\vv{S}_{i,2}\cdot\vv{S}_{i+1,2} \nonumber\\
&+R\sum_{i=1}^{L}\vv{S}_{i,1}\cdot\vv{S}_{i,2},
\label{Hamiltonian}
\end{align}
where $\vv{S}_{i,a}$ is a spin-1/2 operator. The lattice structure is shown in Fig. \ref{fig:lattice}. For $J_1=J_2$, the rung-dimer state is the exact ground state down to a finite critical value of $R$ as shown by Tsukano and Takahashi\cite{tt}. Later, a similar model with a ferromagnetic $J_1$, an antiferromagnetic $J_2$, and an anisotropic ferromagnetic $R$ was investigated by Tonegawa {\it et al.}\cite{tonegawa} Among the variety of ground-state phases of this model, they also found a partial ferrimagnetic phase. In the present work, we consider the whole parameter region with a ferromagnetic $J_1$, an antiferromagnetic $J_2$, and an {\it antiferromagnetic} $R$ without anisotropy. In the remainder of this paper, we set the energy unit by $J_2=1$.

\section{Ground-State Phase Diagram}
\subsection{Numerical analysis}
\begin{figure}[htbp]
\centering{\includegraphics[width=6cm]{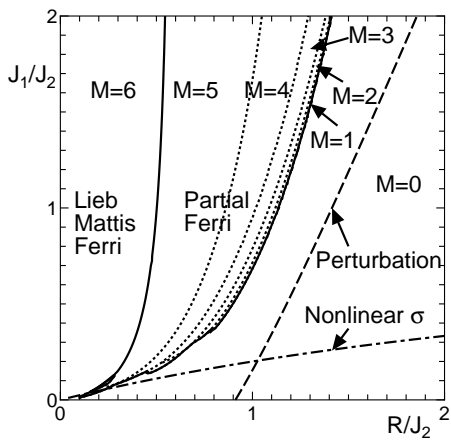}}
\caption{Ground-state phase diagram of the $S=1/2$ Heisenberg ladder (\ref{Hamiltonian}) with $L=12$. The spontaneous magnetization is denoted by $M$. The solid curves are the boundaries of the partial ferrimagnetic phase. The dotted curves are the boundaries between partial ferrimagnetic phases with different magnetization. 
The dashed and dash-dotted lines are the nonmagnetic-partial-ferrimagnetic phase boundaries calculated by the perturbation expansion from the strong-rung limit and the mapping onto the nonlinear $\sigma$ model, respectively.}
\label{fig:phase2}
\end{figure}

The ground-state phase diagram is determined by Lanczos numerical diagonalization with the periodic boundary condition for $L=12$ as shown in Fig. \ref{fig:phase2}.  In the LM ferrimagnetic phase, $M=L/2=6$. In the partial ferrimagnetic phase, $0 < M < L/2$. It is found that the partial ferrimagnetic phase extends over a wide parameter range.

\begin{figure}[htbp]
\centering{\includegraphics[width=60mm]{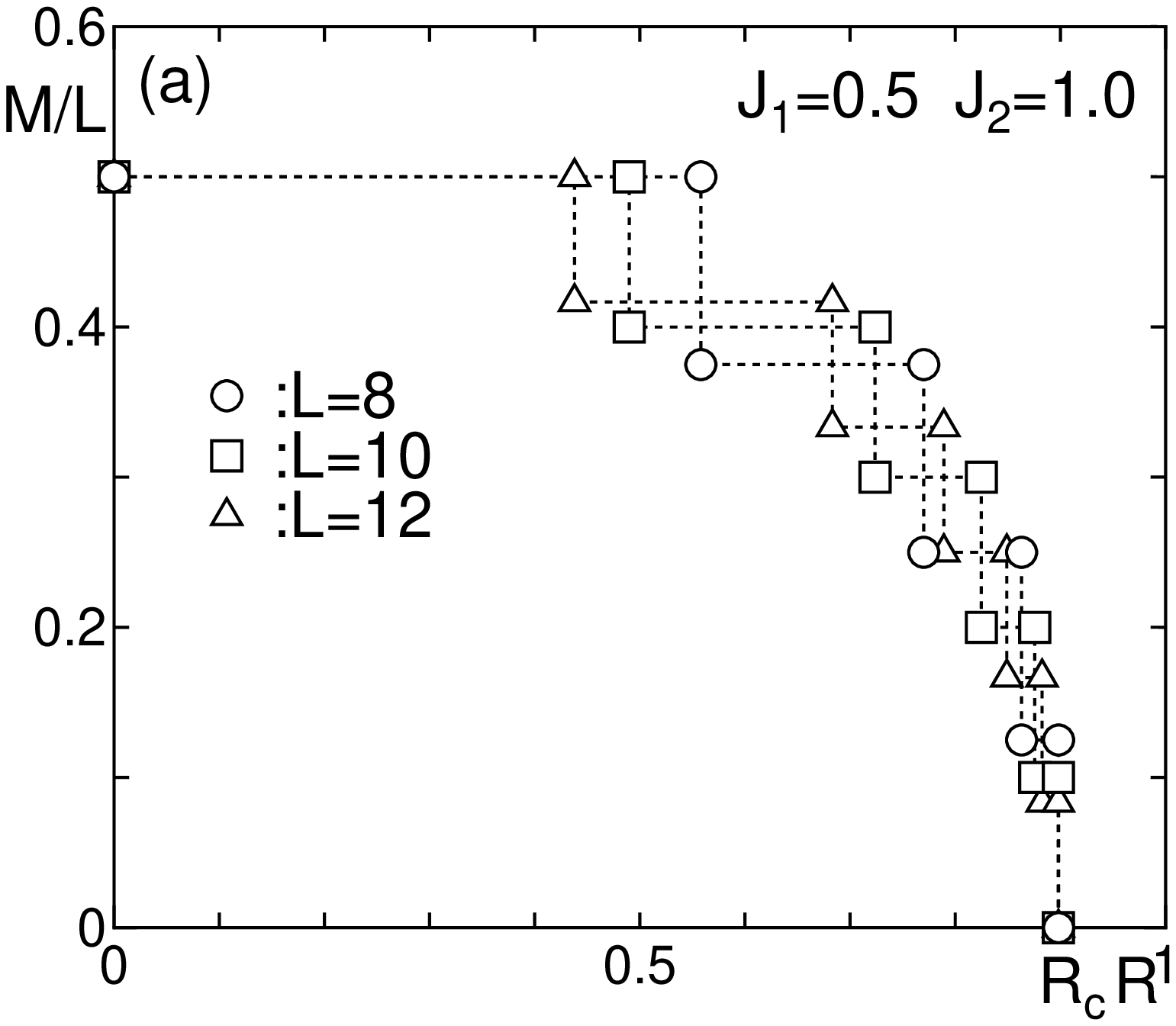}}
\centering{\includegraphics[width=60mm]{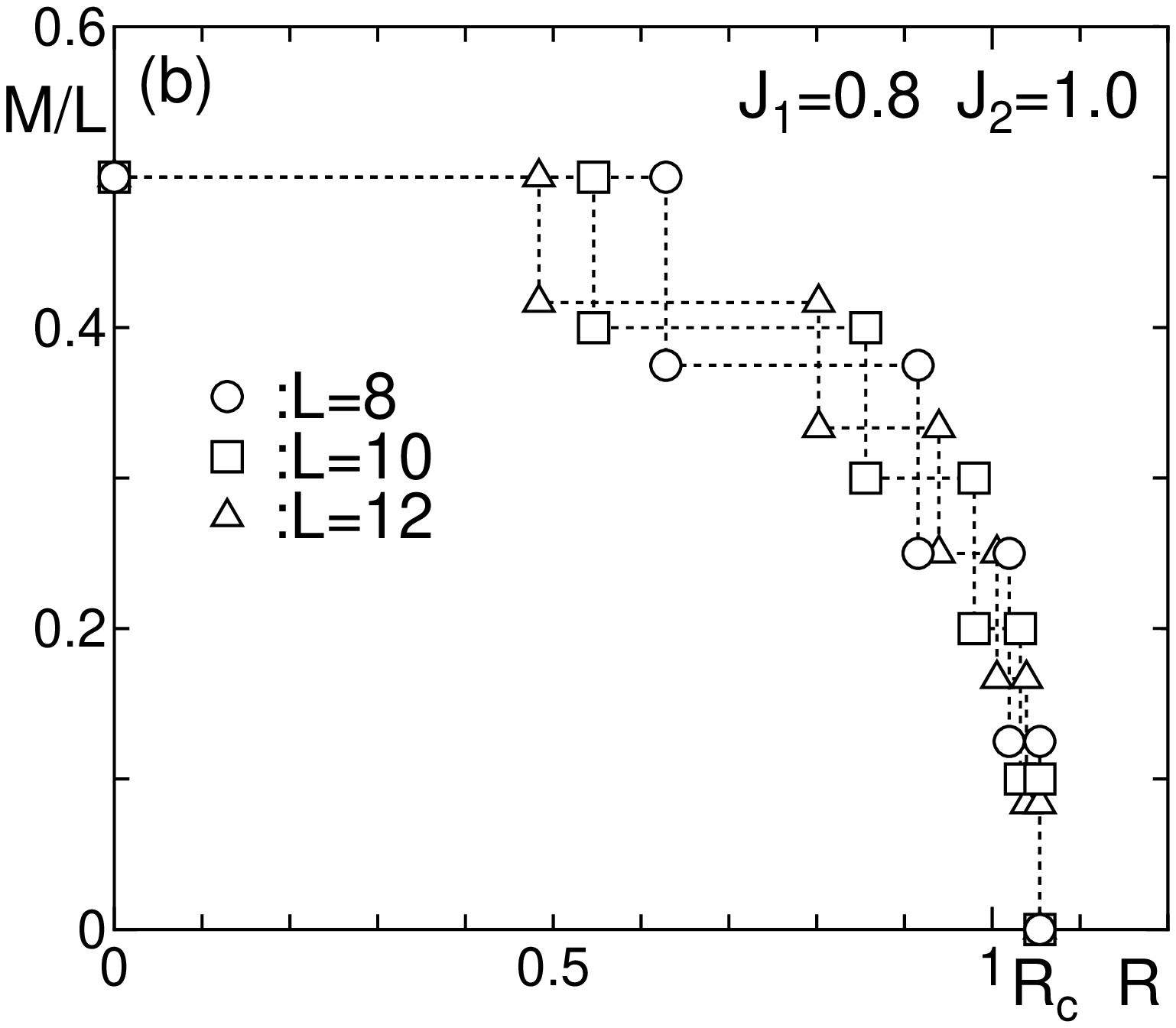}}
\centering{\includegraphics[width=60mm]{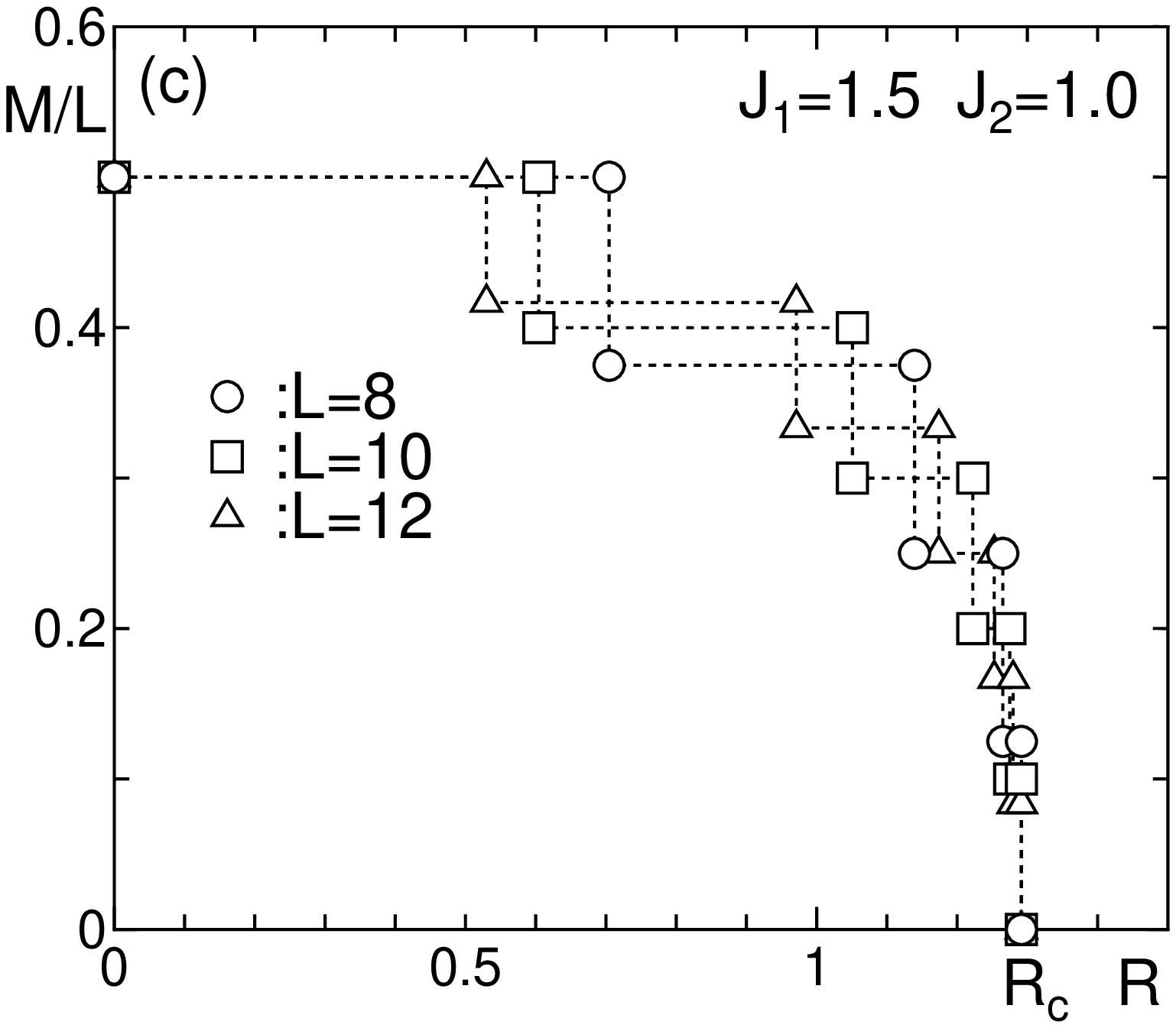}}
\caption{Spontaneous magnetization for (a) $J_1=0.5$, (b) $J_1=0.8$, and (c) $J_1=1.5$.}
\label{fig:mag}
\end{figure}

The $R$-dependences of 
$M$ for $J_1=0.5$, $0.8$, and $1.5$ are presented in Figs. \ref{fig:mag}(a)-\ref{fig:mag}(c), respectively. The critical value $R_{\rm c}$ between the nonmagnetic phase and partial ferrimagnetic phase is insensitive to the system size $L$. For $J_1=0.5$, $0.8$, and $1.5$, we obtain $R_{\rm c}=0.898$, 1.054, and 1.291, respectively.

\begin{figure}[htbp]
\centering{\includegraphics[width=60mm]{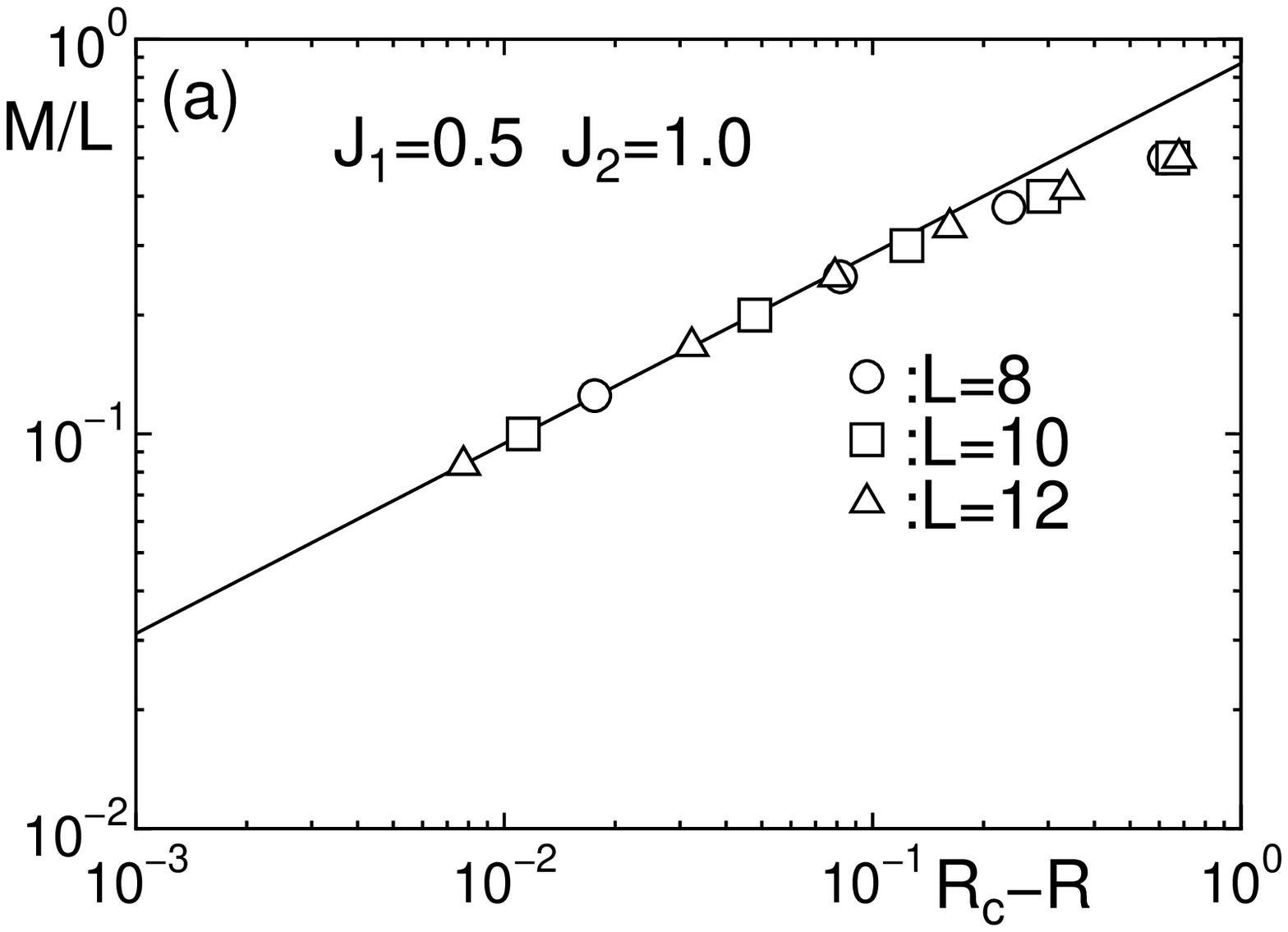}}
\centering{\includegraphics[width=60mm]{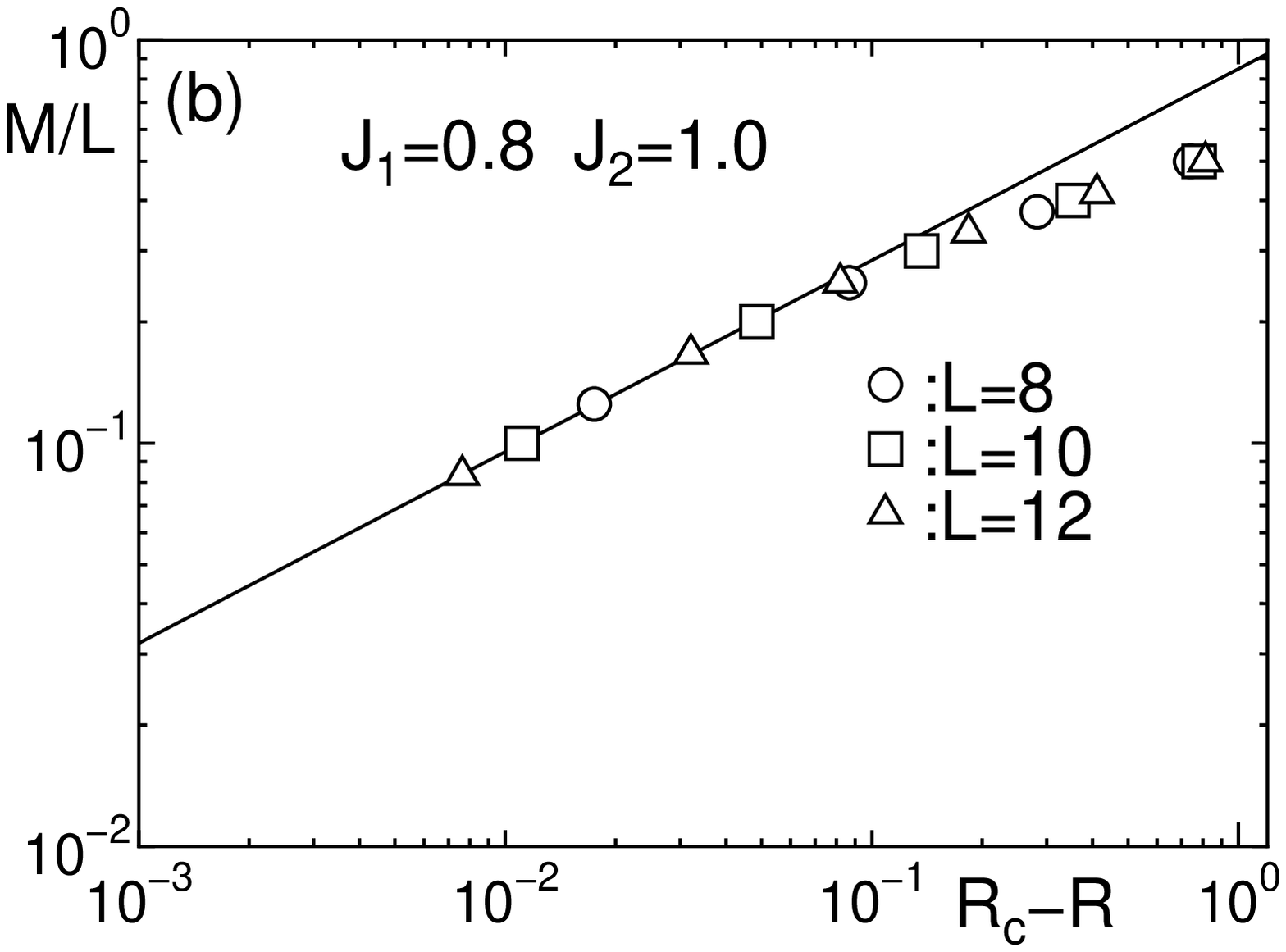}}
\centering{\includegraphics[width=60mm]{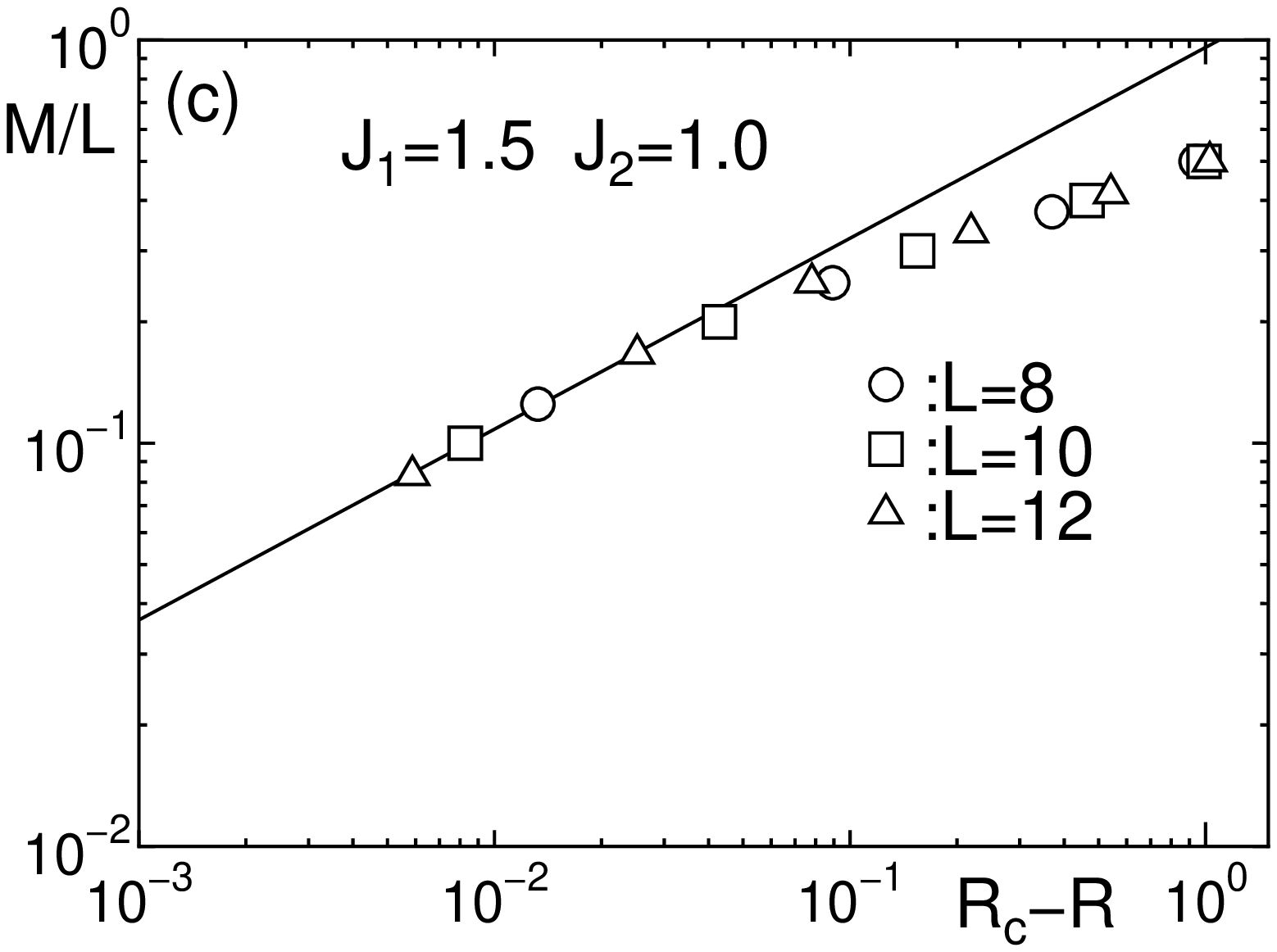}}
\caption{Log-log plot of $M/L$ against $R_{\rm c}-R$ for  (a) $J_1=0.5$, (b) $J_1=0.8$, and (c) $J_1=1.5$.}
\label{fig:mag_log}
\end{figure}

To determine the $R$-dependence of $M$ more precisely,  log-log plots of $M/L$ against $R_{\rm c}-R$ are shown in Fig. \ref{fig:mag_log}. The value of $R$ corresponding to each value of $M/L$ is at the middlepoint of the steps in Fig. \ref{fig:mag}. The solid lines are  fit assuming the form
\begin{align}
\frac{M}{L}=A(R_{\rm c}-R)^{\beta}.
\end{align}
For   $J_1=0.5$, $0.8$, and 1.5, we obtain $\beta=0.48\pm0.01$, $0.48\pm0.01$, and $0.49\pm0.03$, respectively. For $J_1=0.5$ and 0.8, we use two to five points for the fitting. For $J_1=1.5$, we use two to four points. 
 The errors are estimated from the variation of $\beta$ for different choices of the points.
These results are consistent with the estimation of $\beta=1/2$ obtained by a mapping onto the nonlinear $\sigma$ model described in the following subsection.

\begin{figure}[htbp]
\centering{\includegraphics[width=60mm]{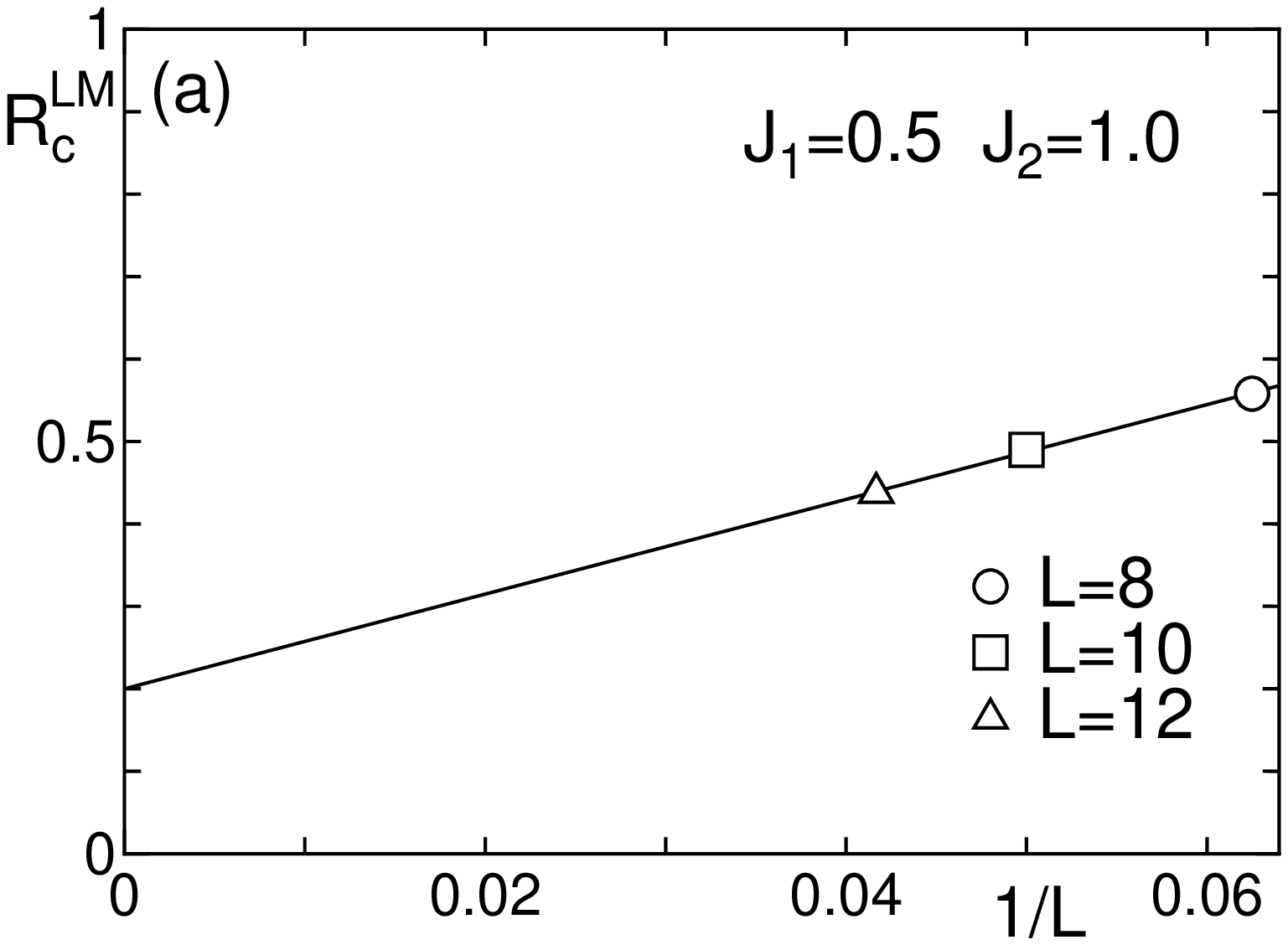}}
\centering{\includegraphics[width=60mm]{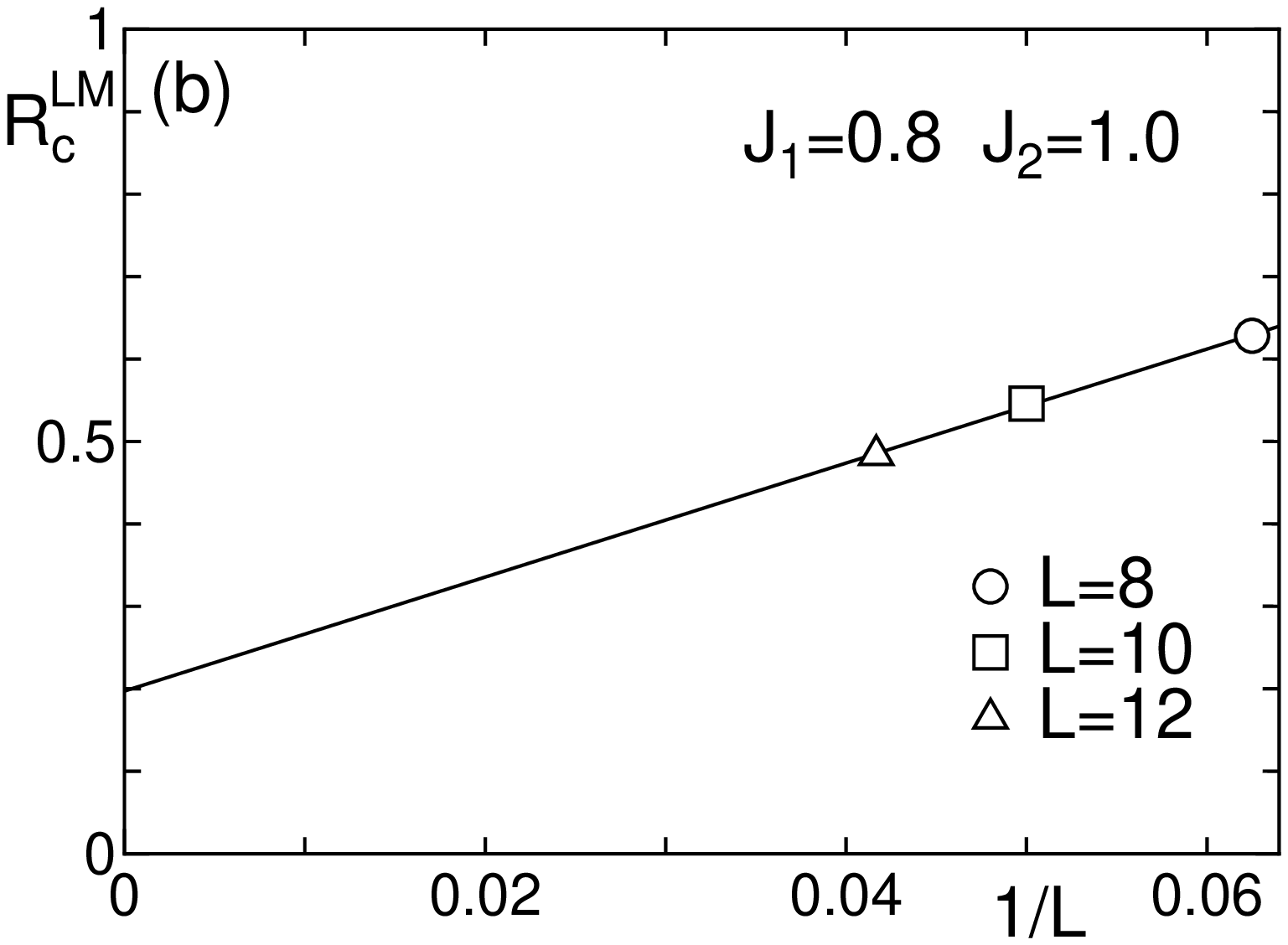}}
\centering{\includegraphics[width=60mm]{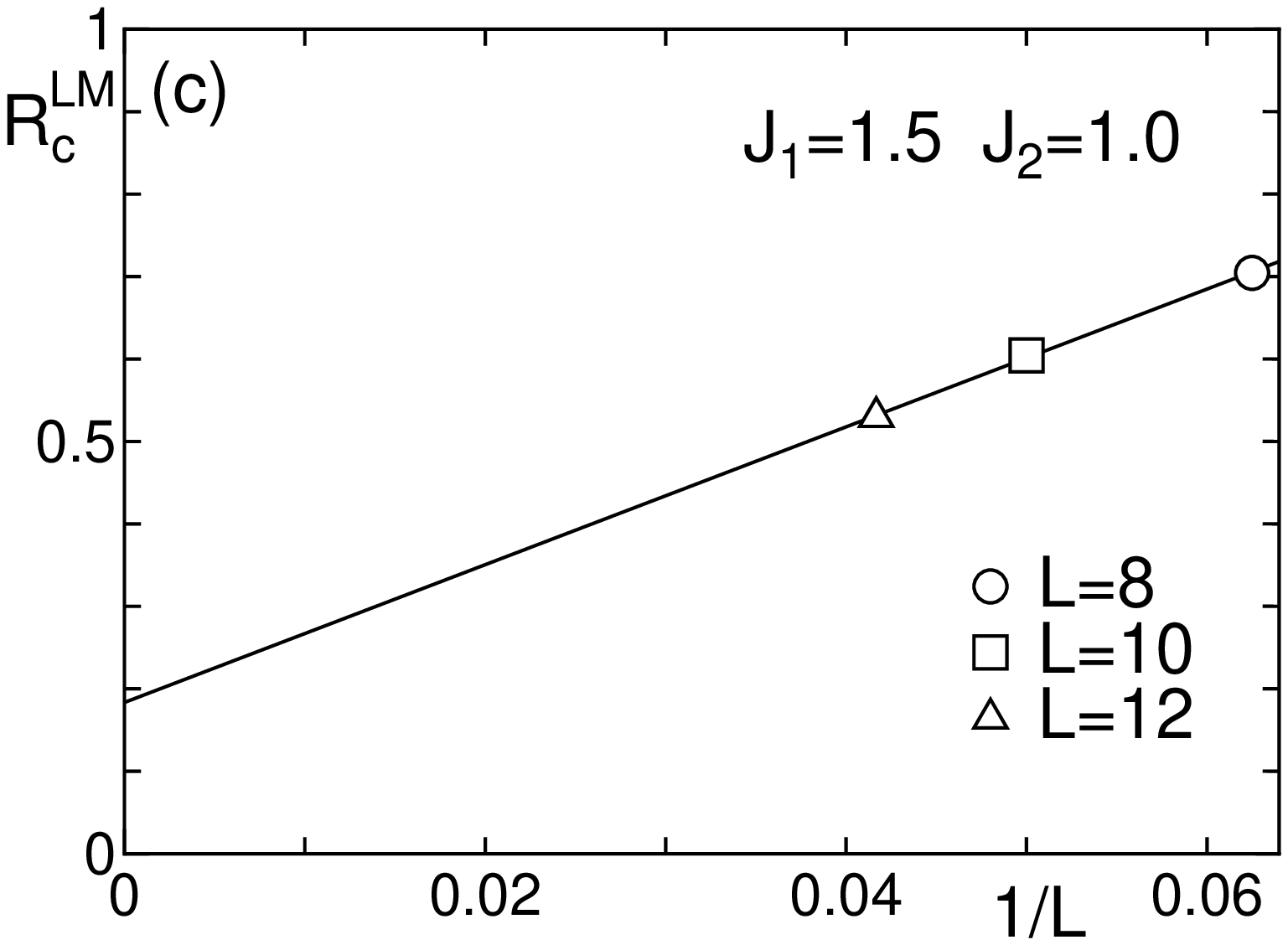}}
\caption{Size dependence of $R_{\rm c}^{\rm LM}$ for (a) $J_1=0.5$, (b) $J_1=0.8$, and $J_1=1.5$.}
\label{fig:mag_l}
\end{figure}

On the other hand, the critical value $R_{\rm c}^{\rm LM}$ between the LM ferrimagnetic phase and the partial ferrimagnetic phase depends strongly on the system size as shown in Fig. \ref{fig:mag}. The size dependences of $R_{\rm c}^{\rm LM}$ are shown in Figs. \ref{fig:mag_l}(a)-\ref{fig:mag_l}(c). The size extrapolation is  carried out using the data for $L=8,10$, and 12. It is noteworthy that $R_{\rm c}^{\rm LM}$ decreases substantially with increasing $L$. The extrapolation suggests  that the LM ferrimagnetic phase is much narrower than that shown in Fig. \ref{fig:phase2} and might eventually vanish.

\subsection{Mapping onto the nonlinear $\sigma$ model}

\subsubsection{Transformation of spin variables}
The ground-state phase diagram is studied analytically by  mapping the Hamiltonian (\ref{Hamiltonian})  onto the nonlinear $\sigma$ model\cite{furuya}. For small $J_1$, the classical ground-state spin configuration of the Hamiltonian (\ref{Hamiltonian}) is given by the N\'eel state
\begin{align}
\vv{S}^{\rm cl}_{i,a}=(-1)^{i+a}S\vv{e}_z.
\end{align}

\begin{figure}[htbp]
\centering{\includegraphics[width=60mm]{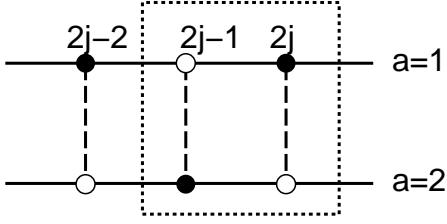}}
\caption{Definition of two sublattices (open and filled circles) and a unit cell (enclosed by a dotted square).}\label{unit-cell}
\end{figure}

Hence, we decompose the whole ladder into two interpenetrating sublattices as shown in Fig. \ref{unit-cell}. Since the unit cell is doubled, we take a unit cell as shown by the dotted square.

We introduce the low-energy modes corresponding to the uniform and staggered components of spin variables $\vv{l}(x_j)$ and $\vv{n}(x_j)$ by\cite{sachdev_book,sierra,sierra_book}
\begin{align}
\so{i}{a}&=\sre{j}{a}{+}\label{eq:soeq1}\\
& \mbox{for}\  (i,a)=(2j-1,1),(2j,2), \notag\\
\so{i}{a}&=\sre{j}{a}{-}\label{eq:soeq2}\\
& \mbox{for}\ (i,a)=(2j-1,2),(2j,1),\notag
\end{align}
which satisfy the constraint
\begin{align}
\vv{n}(x_j)^2=1, \ \ 
\vv{l}(x_j)\cdot\vv{n}(x_j)=0,
\end{align}
where $x_j=2ja_0$ is the coordinate of the center of the $j$th unit cell along the leg.  
 The square root factor $\sqrt{1-A^2_{a}\dfrac{\vv{l}(x_{j})^2}{S^2}}$ is introduced to explicitly normalize $\vv{S}_{i,a}^2$ as
\begin{align}
\vv{S}_{i,a}^2=S^2.
\end{align}
The coefficients $A_a$ are normalized as
\begin{align}
\sum_{a=1}^2A_a=a_0, \label{eq:norm}
\end{align}
so that $\vv{l}(x_j)$ corresponds to the net magnetization per unit cell as
\begin{align}
\vv{l}(x_j)=\frac{1}{2a_0}(\vv{S}_{2j-1,1}+\vv{S}_{2j-1,2}+\vv{S}_{2j,1}+\vv{S}_{2j,2}).
\end{align}

\subsubsection{Stability of the nonmagnetic state}
Taking the continuum limit and within the second order in $\vv{n}'(x_j)$ and $\vv{l}(x_j)$, the Hamiltonian is rewritten as
\begin{align}
\mathcal{H}&=\int\frac{dx}{2a_0}\left(\sum_{a,b=1}^2M_{a,b}A_{a}A_{b}\right)\vv{l}(x)^2  \notag  \\
&-2Sa_0\int\frac{dx}{2a_0}\left(\sum_{a=1}^2J_a A_{a}\right)\nonumber\\
&\times[\vv{l}(x)\cdot\snpo{a}+\snpo{a}\cdot\vv{l}(x)]  \notag  \\
&+2a_0^2S^2\int\frac{dx}{2a_0}\sum_{a=1}^2(-1)^aJ_a(\snpo{a})^2,\label{NLSM-massive}
\end{align}
where
\begin{align}
M=
\begin{pmatrix}
-4J_1+R & R  \\
R & 4J_2+R
\end{pmatrix}.\label{M-def}
\end{align}

To determine $A_a$, we follow Sierra\cite{sierra,sierra_book} to obtain
\begin{align}
A_1=a_0\frac{J_2}{J_2-J_1}~~,~~A_2=a_0\frac{-J_1}{J_2-J_1}.\label{eq:A-def}
\end{align}
Using Eq. (\ref{M-def}) and Eq. (\ref{eq:A-def}), we have
\begin{align}
\sum_{a,b=1}^2M_{a,b}A_aA_b=a_0^2\frac{(J_2-J_1)R-4J_1J_2}{J_2-J_1}. 
\end{align}
Hence, we finally obtain
\begin{align}
\mathcal{H}&=\int\frac{dx}{2a_0}2a_0^2S^2(J_2-J_1)\vv{n}'(x)^2  \notag  \\
&+\int\frac{dx}{2a_0}\left[a_0^2\frac{(J_2-J_1)R-4J_1J_2}{J_2-J_1}\right] \vv{l}(x)^2. 
  \label{NLSM-lad}
\end{align}
Limiting ourselves to the case of $J_2 \gg J_1$, the state with $\vv{l}(x)=0$ is unstable for $R < R_{\rm c}$, where
\begin{align}
R_{\rm c}=\frac{4J_1J_2}{J_2-J_1}. \label{eq:rc}
\end{align}
The instability in $\vv{l}$ implies the transition to the ferrimagnetic state.  The critical value given by Eq. (\ref{eq:rc}) is plotted in  Fig. \ref{fig:phase2} by a dash-dotted line. Considering that the present approximation is valid for $J_2 \ll J_1$, it is consistent with the phase boundary obtained by numerical calculation.

\subsubsection{Higher-order correction in $\vv{l}(x)$}

We have to consider the higher-order correction in  $\vv{l}(x)$ to fix the equilibrium value of  $\vv{l}(x)$ in the unstable region $R > R_{\rm c}$. Hence, we expand Eq. (\ref{eq:soeq1}) and Eq. (\ref{eq:soeq2}) up to $O(\vv{l}^4)$. Then, the Hamiltonian yields
\begin{align}
\mathcal{H}&=\int\frac{dx}{2a_0}2a_0^2S^2(J_2-J_1)\vv{n}'(x)^2  \notag  \\
&+\int\frac{dx}{2a_0}\left[a_0^2\frac{(J_2-J_1)R-4J_1J_2}{J_2-J_1}\vv{l}(x)^2\right]\\
&+\int\frac{dx}{2a_0}\frac{a_0^4}{4S^2}\left(\frac{J_2+J_1}{J_2-J_1}\right)^2R[\vv{l}(x)^2]^2.  \label{NLSM-lad_4}
\end{align}

The coefficient of $[\vv{l}(x)^2]^2$ is positive definite. Hence, the magnitude of the equilibrium value of $\vv{l}$ grows continuously from $R = R_{\rm c}$ as
\begin{align}
|\aver{\vv{l}}|=\frac{\sqrt{2}J_1J_2S}{R_c(J_1+J_2)a_0}\sqrt{\frac{R_c-R}{R}}\propto S\sqrt{R_c-R}.
\end{align}
The magnitude of the uniform magnetization per site ${M}$ is given by
\begin{align}
{M}=\frac{|\aver{\vv{l}}|a_0}{2}.
\end{align}
This result implies $\beta=1/2$ as estimated numerically.

 In the higher-order terms, the terms such as $\vv{l}(x)^2\vv{n}'(x)^2$ and $\vv{l}(x)^2[\vv{l}(x)\cdot\vv{n}'(x)+\vv{n}'(x)\cdot\vv{l}(x)]$ also appear. Replacing $\vv{l}(x)^2$ by $\aver{\vv{l}(x)}^2$, the first term is absorbed by a slight redefinition of the coefficient of $\vv{n}'(x)^2$ and the second term leads to a small but finite topological angle. In the magnetized sector, however, the ground state is an SMTLL, as discussed below, and the topological angle does not play an essential role. The term $\vv{l}'(x)^2$ also appears with a positive coefficient. This term suppresses the spatial variation of $\vv{l}(x)$ and stabilizes the ferrimagnetic long-range order. 
Hence, we conclude that a second-order transition to a partial ferrimagnetic phase takes place for $R < R_{\rm c}$. 

\begin{figure}[htbp]
\centering{\includegraphics[width=7cm]{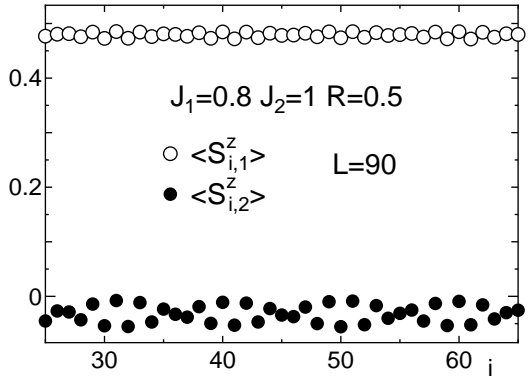}}
\caption{Ground-state expectation values $\aver{S^z_{i,a}} (a=1,2)$ for $J_1=0.8, J_2=1$, and $R=0.5$ with $L=90$ near the center of the whole ladder.}
\label{fig:szav}
\end{figure}

Following the argument of Ref. \citen{furuya}, this ground state is an SMTLL with broken SU(2) symmetry down to U(1). Hence, the incommensurate quasi-long-range modulation of magnetization is also expected in the partial ferrimagnetic phase. This is confirmed by the finite-size DMRG calculation of the expectation values $\aver{S^z_{i,a}} (a=1,2)$ as shown in Fig. \ref{fig:szav} for $J_1=0.8, J_2=1,$ and $R=0.5$ with system size $L=90$. Similar behavior is also found for several other values of the parameters within  the partial ferrimagnetic phase. In each DMRG step, the number $m$ of states  kept in each subsystem is 240. The convergence with respect to $m$ is confirmed. Although a true breakdown of the translational symmetry is absent in the infinite SMTLL state, the oscillatory modulation of magnetization becomes visible in spin expectation values $\aver{S^z_{i,a}}$ owing to the presence of open boundaries. 

\subsection{Perturbation from the strong-rung limit}

In the strong-rung limit $R \gg J_1, J_2$, we divide the Hamiltonian (\ref{Hamiltonian}) as
\begin{align}
\mathcal{H}&=\mathcal{H}_0+\mathcal{H}_1,  \\
\mathcal{H}_0&=R\sum_{i=1}^L\vv{S}_{i,1}\cdot\vv{S}_{i,2},  \\
\mathcal{H}_1&=-J_1\sum_{i=1}^{L}\vv{S}_{i,1}\cdot\vv{S}_{i+1,1}+J_2\sum_{i=1}^{L}\vv{S}_{i,2}\cdot\vv{S}_{i+1,2}.
\end{align}
In this subsection, we regard $\mathcal{H}_0$ as an unperturbed Hamiltonian and $\mathcal{H}_1$ as a perturbation Hamiltonian.

Each spin state is described by eigenstates of $S_{i,a}^z$ as
\begin{align}
\ket{\uparrow_{i,a}}=\left|S_{i,a}^z=\frac{1}{2}\right\rangle,~~\ket{\downarrow_{i,a}}=\left|S_{i,a}^z=-\frac{1}{2}\right\rangle.
\end{align}
The singlet and triplet states on each rung are defined by
\begin{align}
|s_i\rangle&=\frac{1}{\sqrt{2}}(|\uparrow_{i,1}\rangle|\downarrow_{i,2}\rangle-|\downarrow_{i,1}\rangle|\uparrow_{i,2}\rangle),  \\
|t_i^+\rangle&=|\uparrow_{i,1}\rangle|\uparrow_{i,2}\rangle,  \\
|t_i^0\rangle&=\frac{1}{\sqrt{2}}(|\uparrow_{i,1}\rangle|\downarrow_{i,2}\rangle+|\downarrow_{i,1}\rangle|\uparrow_{i,2}\rangle),  \\
|t_i^-\rangle&=|\downarrow_{i,1}\rangle|\downarrow_{i,2}\rangle.
\end{align}

In the limit $R\rightarrow\infty$, the ground state is the rung singlet (RS) state $|{\rm RS}\rangle$ defined by
\begin{align}
|{\rm RS}\rangle&=|s_1\rangle|s_2\rangle\cdots|s_L\rangle.
\end{align}
This is an eigenstate of $\mathcal{H}_0$ that satisfies
\begin{align}
\mathcal{H}_0|{\rm RS}\rangle=-\frac{3}{4}RL|{\rm RS}\rangle .
\end{align}
The eigenvalue of this RS state is 
\begin{align}
E_{\rm RS}=-\frac{3}{4}RL-\frac{3}{32}\frac{(J_1-J_2)^2}{R}L+\mathcal{O}(R^{-2})
\end{align}
up to the second order in $J_1$ and $J_2$.

In the single triplet (RT1) state, one of the rung singlets is replaced by a rung triplet. Owing to the translational invariance, the eigenstate is a plane-wave state indexed by a wave number $k$ ($-\pi/a_0\leq k\leq\pi/a_0$) and $\alpha\ (=0,\pm)$  as  
\begin{align}
|{\rm RT1};k,\alpha\rangle=\frac{1}{\sqrt{L}}\sum_{i=1}^L\exp(ikx_i)|s_1\rangle|s_2\rangle\cdots|t_i^{\alpha}\rangle\cdots|s_L\rangle.  \label{RT1state}
\end{align}
This is an eigenstate of $\mathcal{H}_0$ that satisfies
\begin{align}
\mathcal{H}_0\ket{{\rm RT1};k,\alpha}
&=\left(-\frac{3}{4}RL+R\right)\ket{{\rm RT1};k,\alpha}.
\end{align}
Up to the second-order perturbation in $J_1$ and $J_2$, the eigenvalue of $\ket{{\rm RT1};k,\alpha}$ is given by
\begin{align}
E_{\rm RT1}^\alpha(k)&=-\dfrac{3}{4}RL-\frac{3}{32}\frac{(J_1-J_2)^2}{R}L+R-\frac{J_1J_2}{R}  \notag  \\
&+\left[\dfrac{J_2-J_1}{2}+\frac{(J_1+J_2)^2}{4R}\right]\cos k\notag\\
&-\frac{(J_1-J_2)^2}{8R}\cos^2k+\mathcal{O}(R^{-2}).
\end{align}
The minimum of $E_{\rm RT1}$ located at $k=0$ or $\pi$ is given by
\begin{align}
E^{\rm min}_{\rm RT1}&=E_{\rm RS}+R-\frac{J_1J_2}{R}-\left|\dfrac{J_2-J_1}{2}+\dfrac{(J_1+J_2)^2}{4R}\right|\notag\\
&-\frac{(J_1-J_2)^2}{8R}+\mathcal{O}(R^{-2}).
\end{align}
Hence, if 
\begin{align}
R-\frac{J_1J_2}{R}-\left|\dfrac{J_2-J_1}{2}+\dfrac{(J_1+J_2)^2}{4R}\right|-\frac{(J_1-J_2)^2}{8R}<0  \label{ineqR}
\end{align}
is satisfied, the RS state is unstable against the formation of a rung-triplet excitation. This instability leads to ferrimagnetic ordering. The critical value of $R$ is given by
\begin{align}
R_{\rm c}=\frac{1}{4}(J_2-J_1+\sqrt{7J_1^2+18J_1J_2+7J_2^2}).
\end{align}
This is plotted in  Fig. \ref{fig:phase2} by a dashed line. Considering that the present approximation is valid for $R \gg J_1, J_2$, it is qualitatively consistent with the phase boundary obtained by numerical calculation. This expression reduces to $R_{\rm c}=\sqrt{2}$ obtained by Tsukano and Takahashi\cite{tt} for $J_1=J_2=1$.

\section{Finite-Temperature Properties}
The ground state of the present system is an SMTLL, analogous to the ground state of a spin chain in the effective magnetic field. Nevertheless, the ferromagnetic long-range order is destroyed at finite temperatures due to one-dimensionality. This implies that the effective magnetic field vanishes as soon as the temperature becomes finite. Hence, the finite-temperature properties of the present system are not simply described as those of a conventional Tomonaga--Luttinger liquid (TLL) at finite temperatures. This situation poses the nontrivial question ``What is the fate of the SMTLL at finite temperatures?''.

\begin{figure}[htbp]
\centering{\includegraphics[width=6cm]{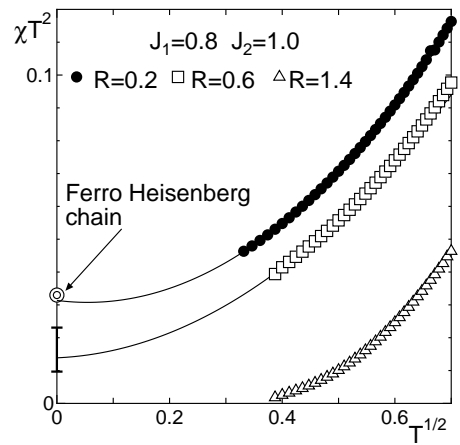}}
\caption{Plot of $\chi T^2$ against $T^{1/2}$. The solid curves are fit by Eq. (\ref{eq:cht2}). }
\label{fig:sus}
\end{figure}

To obtain insight into this question, the  finite-temperature susceptibility is calculated by the cTPQ method.\cite{ss,hams-deraedt,imada-takahashi,jp} The average is taken over 1200 initial vectors. The size extrapolation is carried out by the Shanks transform\cite{shanks} from $L=8, 10$, and 12. Motivated by the low-temperature behavior of the susceptibility of the $S=1/2$ ferromagnetic Heisenberg chain,\cite{ty} we fit the data by the formula
\begin{align}
\chi T^2\simeq C_0+C_1T^{1/2}+C_2T.\label{eq:cht2}
\end{align}
The plot of $\chi T^{2}$ against $T^{1/2}$ is shown in Fig. \ref{fig:sus}. This plot suggests that $C_0 >0$  in the partial ferrimagnetic phase as well as in the LM ferrimagnetic phase. This means that the susceptibility in these phases behaves as $\chi \sim T^{-2}$ at low temperatures.

Although the above result is not conclusive due to the limited system size, the following physical argument supports the validity of this behavior. In addition to the excitations of the conventional TLL, whose excitation energy is normally proportional to the wave number $k$,  the ferromagnetic fluctuation modes coexist as low-lying modes in the SMTLL. The amplitude of the ferromagnetic fluctuation mode in the long-wave-length limit is simply the total magnetization, which commutes with the Hamiltonian. Therefore, similarly to the ferromagnetic fluctuations in the one-dimensional ferromagnets, their excitation energy is proportional to $k^2$. Hence, the time scale of the ferromagnetic fluctuation modes is much longer than that of the excitations in the conventional TLL for small $k$. This implies that the whole system can be regarded as a TLL in the background of slowly fluctuating almost uniform ferromagnetic modes. The latter modes contribute to the finite-temperature susceptibility in the same way as the ferromagnetic modes do in a ferromagnetic chain, leading to the behavior $\chi \sim T^{-2}$ at low temperatures. 

\section{Summary and Discussion}
We have investigated the ground-state properties of $S=1/2$ Heisenberg ladders with a ferromagnetic leg, an antiferromagnetic leg, and antiferromagnetic rungs using Lanczos diagonalization. It is shown that a partial ferrimagnetic phase extends over a wide parameter range. The numerical results are supported by analytical calculations using the nonlinear $\sigma$ model and the perturbation expansion from the strong-rung limit. 

The finite-temperature magnetic susceptibility is calculated using the cTPQ method. Although the ground state is an SMTLL, the finite-temperature properties are expected to be different from those of a conventional TLL, since the spontaneous magnetization vanishes at finite temperatures. Our numerical results suggest that the susceptibility diverges as $T^{-2}$ in the ferrimagnetic phases as in the case of a ferromagnetic Heisenberg chain. 

This behavior can be understood if we regard the present system as a TLL in a slowly fluctuating ferromagnetic background. Since the ferromagnetic spin wave  has much lower excitation energy than the TLL excitation in the long-wavelength limit, this should make a dominant contribution to the susceptibility.  Nevertheless, the details of the properties of  the SMTLL at finite temperatures still remain to be investigated. It is hoped that extensive analyses of other models with ground states of this kind will clarify their generic nature.

\acknowledgments

The authors are grateful to S. C. Furuya for enlightening comments and discussion on the nonlinear $\sigma$ model analysis. They thank the authors of Ref. \citen{tonegawa} for the discussion and showing their results prior to their publication. They also thank H. Shinaoka and K. Yoshimi for  advice on the cTPQ method. For the numerical diagonalization, the package TITPACK ver. 2 coded by H. Nishimori was used.  Part of the numerical computation in this work was carried out using the facilities of the Supercomputer Center, Institute for Solid State Physics, University of Tokyo, and   Yukawa Institute Computer Facility in Kyoto University. This work was supported by JSPS KAKENHI Grant Number JP25400389.

\end{document}